\documentclass[12pt,a4]{article}
\usepackage{epsfig}
\usepackage{amssymb,latexsym,amsmath}

\begin{document}
\begin{center}
{\Large \bf Particle Number Fluctuations\\

\vspace{0.3cm}
in Canonical Ensemble}
\end{center}

\vspace{0.3cm}
\begin{center}
{\bf V.V. Begun}$^{a}$, {\bf M. Ga\'zdzicki}$^{b,c}$, {\bf M.I.
Gorenstein}$^{a,d}$ and {\bf O.S. Zozulya}$^{d,e}$

\vspace{1cm}

\noindent
\begin{minipage}[t]{12.5cm}
$^{a}$ Bogolyubov Institute for Theoretical Physics, Kiev, Ukraine\\
$^b$ Institut f\"ur Kernphysik, Universit\"at Frankfurt, Germany\\
$^c$ \'Swi\c{e}tokrzyska Academy, Kielce, Poland \\
$^{d}$ Institut f\"ur Theoretische Physik, Universit\"at Frankfurt,
Germany\\
$^e$ Taras
Shevchenko Kiev National University, Kiev, Ukraine\\
\end{minipage}
\end{center}

\begin{abstract}
Fluctuations of charged particle number are
studied
in the canonical ensemble. 
In the infinite volume limit 
the fluctuations in the canonical ensemble are different from the
fluctuations in the grand canonical one.
Thus, the well-known equivalence of both ensembles for the average
quantities does not extend for the fluctuations.
In view of a possible  relevance of the results
for the analysis of fluctuations
in nuclear collisions at high energies, a role of the
limited kinematical acceptance is studied.
\end{abstract}

\vspace{1.0cm} \noindent {\bf 1. Introduction.}~~ 
The statistical approach 
to strong interactions is surprisingly 
successful in describing experimental results
on hadron production properties in nuclear collisions at high energies
(see  e.g. Ref.~\cite{PBM} and references therein). 
This motivates a rapid development of statistical models and it raises
new questions, previously not addressed in
statistical physics. 
In particular, an applicability of the models formulated within
various statistical ensembles has been considered for  average 
quantities.
The micro-canonical ensemble, where the motional and material conservation
laws are strictly fulfilled in all microscopic states of the system,
has to be used for collisions in which a small number of particles is
produced, like p+p interactions at low energies (see
e.g. Ref.~\cite{mce}). 
The canonical ensemble ($~c.e.~$), where only the material conservation
laws are obeyed, is relevant for the systems with a large number of
all produced particles, but a small number of carriers of conserved
charges 
like electric charge, baryon number, strangeness or charm 
(see e.g. Ref.~\cite{ce}).
Finally, models formulated using grand canonical
ensemble ($~g.c.e.~$)
can be used when the number of carriers of conserved charge is  
large enough (see e.g. Ref.~\cite{gce}).
In the latter approach both material and motional conservation
laws are relaxed and the mean values of conserved charges and energy
are adjusted by introduction of chemical potentials and temperature,
respectively.

The question of applicability of various statistical ensembles
for the study of fluctuations of physical quantities has not been addressed
up to now.
In the text-books of statistical mechanics, the particle number
fluctuations are considered in $g.c.e$ only.
It is because 
the discussion is limited to the 
non-relativistic cases,
so that in the $c.e.$ the particle number is fixed. 
However, in the relativistic case, relevant for the models of hadron
production in high energy nuclear collisions, only conserved
charges are fixed, and consequently the particle number fluctuates
in both $c.e.$ and $g.c.e.$ .

The analysis of fluctuations is an important tool to study a physical
system created in high energy nuclear collisions (see e.g. \cite{fluc}). 
Recently, rich experimental data on fluctuations of particle
production properties in nuclear collisions at high energies
have been presented (see e.g. presentations at ``Quark Matter 2004'').
In particular,
intriguing results concerning the particle number fluctuations
in collisions of small nuclei at the CERN SPS 
have been shown \cite{mgqm}. 
These new results motivate our work, in which the particle number
fluctuations are calculated in $c.e.$ and compared with those obtained
in $g.c.e.$ .
Finally, the possible influence of the limited experimental acceptance on
observed fluctuations is studied.

\vspace{0.3cm} \noindent {\bf 2. Partition function in g.c.e. and
c.e.}~~ Let us consider the system which consists of one sort of
positively and negatively charged particles (e.g. $\;\pi^+\;$ and
$\;\pi^-\;$  mesons) with total charge equal to zero $\;Q=0\;$.
In the case of the Boltzmann ideal gas (the interactions
and quantum statistics effects are neglected) in the volume $V$ and at temperature
$T$ the g.c.e. partition function reads:
\begin{align}\label{Zgce}
Z_{g.c.e.}(V,T) \;=\;
 \sum_{N_+=0}^{\infty}\sum_{N_-=0}^{\infty}\;
 \frac{(\lambda_+z)^{N_+}}{N_+!}\;\frac{(\lambda_-z)^{N_-}}{N_-!}
\;=\; \exp\left(\lambda_{+}z~+~\lambda_{-}z\right)~=~\exp(2z)~.
\end{align}
In Eq.~(\ref{Zgce}) $\;z\;$ is a single particle partition function
\begin{align}\label{z}
z\;=\; \frac{V}{2\pi^2}
       \int_{0}^{\infty}k^{2} dk\;
       \exp\left[-~\frac{(k^{2}+m^{2})^{1/2}}{T}\right]
        \;=\; \frac{V}{2\pi^2} \;\;
       T\,m^2\,K_2\left(\frac{m}{T}\right)~,
\end{align}
where $m$ is a particle mass and $\;K_2\;$ is the modified Hankel
function.
Parameters $\;\lambda_+\;$ and $\;\lambda_-\;$ are auxiliary
parameters introduced  in order to calculate the mean number and
the fluctuations of positively and negatively charged particles (the chemical
potential  equals to zero to satisfy the condition
$\langle Q\rangle_{g.c.e.}=0$). They are set to one in the
final formulas.

The $c.e.$  partition function 
is obtained by an explicit introduction of the charge
conservation constrain, $N_+ - N_- = 0$ for each microscopic state
of the system and it reads: 
\begin{align}\label{Zce}
Z_{c.e.}(V,T)
 &\;=\;
 \sum_{N_+=0}^{\infty}\sum_{N_-=0}^{\infty}\;
 \frac{(\lambda_+ z)^{N_+}}{N_+!}\;\frac{(\lambda_- z)^{N_-}}{N_-!}
 \;\delta (N_+-N_-) \;=\;
  \\
 &\;=\;
 \frac{1}{2\pi}\int_0^{2\pi}d\phi\;\;
   \exp\left[ z\;(\lambda_+\;e^{i\phi}
                   \;+\; \lambda_-\;e^{-i\phi})\right]
  \;=\; I_0(2z)\;. \nonumber
\end{align}
In Eq.~(\ref{Zce})  the integral representations of the
$\delta$-Kronecker symbol
and the modified Bessel function were used:
\cite{I}
\begin{align} \label{IQ}
\delta(n) = \frac{1}{2\pi}\int_0^{2\pi}d\phi~ \exp(in\phi)~,~~
I_Q(2z) =  \frac{1}{2\pi}\int_0^{2\pi}d\phi\;
   \exp[-i Q \phi\;+\;2z\;\cos\phi] \;.
\end{align}

\vspace{0.3cm} \noindent {\bf 3. Mean particle number.}~~ The average
number of $N_{+}$ and $N_{-}$ can be calculated as
\begin{align} \label{average}
\langle N_{\pm}\rangle \;=\;
 \left( \frac{\partial}{\partial\lambda_{\pm}}\ln Z
 \right)_{\lambda_{\pm}\;=\;1}~,
\end{align}
and in the $\;g.c.e.\;$ (\ref{Zgce}) they are  
equal to:

\begin{align} \label{gce-average}
\langle N_{\pm}\rangle_{g.c.e.} \;=\;
 \left( \frac{\partial}{\partial\lambda_{\pm}}\ln Z_{g.c.e.}
 \right)_{\lambda_{\pm}\;=\;1} =\; z~.
\end{align}
By construction the mean total charge is equal to zero:
$\langle
Q\rangle _{g.c.e.}= \langle N_{+} \rangle_{g.c.e.} - \langle N_{-}
\rangle_{g.c.e.} = 0$.
In the c.e. (\ref{Zce}) the charge conservation $Q=N_{+}-N_{-}=0$ is
imposed on each microscopic state of the system. This condition introduces
a  correlation between particles which carry conserved charges.
The average particle numbers are \cite{ce}:
\begin{align}\label{ce-average}
\langle N_{\pm}\rangle_{c.e.} \;=\;
 \left( \frac{\partial}{\partial\lambda_{\pm}}\ln Z_{c.e.}
 \right)_{\lambda_{\pm}\;=\;1}
\;=\; z\;\frac{I_{1}(2z)}{I_0(2z)}\;.
\end{align}
The exact charge conservation leads to the $\;c.e.\;$ suppression
($I_{1}(2z)/I_{0}(2z)<1$) of the charged particle multiplicity relative to
the result for the $\;g.c.e.\;$ (\ref{gce-average}). 
The ratio of $\langle N_{\pm} \rangle$ calculated in the
$~c.e.~$ and $~g.c.e.~$ is plotted as a function of $z$ in Fig.~1.

\vspace{-0.5cm}
\begin{figure}[h!]
 \epsfig{file=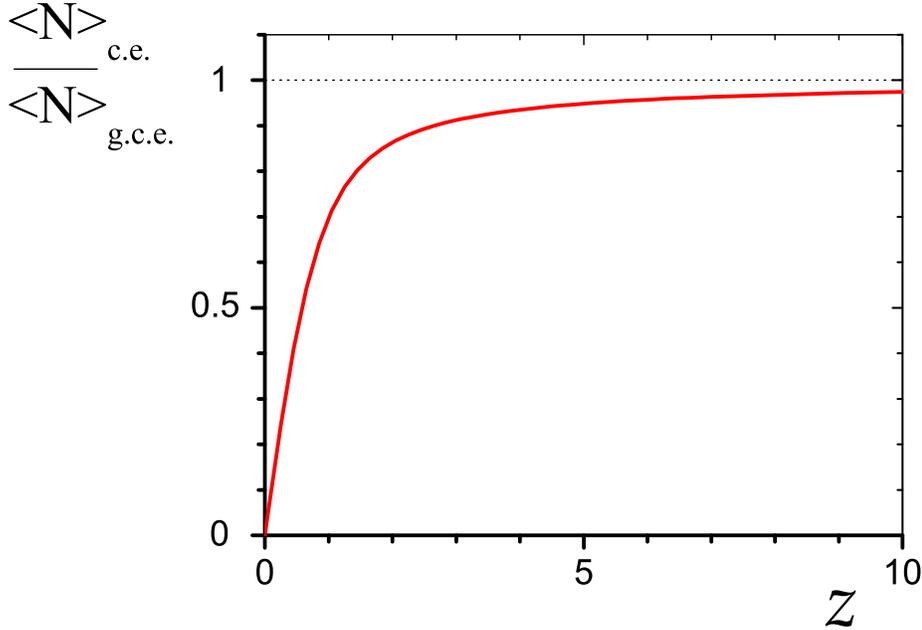,height=10cm,width=14cm}
                                                                                
 \vspace{-1cm}
 \caption{The ratio of $\langle N_{\pm}\rangle_{c.e.}$ (\ref{ce-average})
 to $\langle N_{\pm} \rangle _{g.c.e.}$ (\ref{gce-average})  as a
 function of $z$.
 }\label{fig-Nce/Ngce}
\end{figure}

\noindent
In the
large volume limit ($V\rightarrow \infty$ corresponds also to
$z\rightarrow \infty$) the 
results for mean quantities in the
$\;c.e.\;$ and $\;g.c.e.\;$ are equal.
This result is referred as an equivalence of
the canonical and grand canonical ensembles.
It can be obtained using an  asymptotic expansion
of the modified Bessel function \cite{I}:
\begin{align}\label{Bessel-0lim}
\lim_{z\to\infty}I_n(2z)\;=\;\frac{\exp(2z)}{\sqrt{4\pi z}}\;\left[1~-~
\frac{4n^{2}-1}{16z}~+~O\left(\frac{1}{z^{2}}\right)\right]~,
\end{align}
which gives $I_{1}(2z)/I_{0}(2z) \rightarrow 1$ and therefore
\begin{align}\label{ce1}
\langle N_{\pm} \rangle _{c.e.}~\cong~ \langle N_{\pm} \rangle _{g.c.e} =
z~.
\end{align}
Using the
series expansion one gets \cite{I} for small systems ($z\ll 1$):
\begin{align} \label{Bessel-1lim}
I_{n}(2z)~=~\frac{z^{n}}{n!}~+~\frac{z^{n+2}}{(n+1)!}~
+~O\left(z^{n+4}\right)~,
\end{align}
and consequently  $I_{1}(2z)/I_{0}(2z)\cong z$ which results in
\begin{align}\label{ce2}
\langle N_{\pm}\rangle_{c.e.}~\cong ~z^{2}~~ \ll  ~~ \langle
N_{\pm}\rangle_{g.c.e.}~=~z~.
\end{align}
The asymptotics of the mean multiplicity discussed above 
are clearly seen in Fig.~\ref{fig-Nce/Ngce}.
%
%

\vspace{0.3cm} \noindent {\bf 4. Scaled variance.}~~ An
useful measure of fluctuations of any
variable $X$ is the ratio of its variance $V(X)=\langle X^2\rangle
-\langle X \rangle ^2$ to its mean value $\langle X \rangle$, 
referred here as the  
scaled variance:
\begin{align} \label{omega}
 \omega^{X} \;\equiv \; \frac{\langle X^2\rangle - \langle
X\rangle^2}{\langle
X\rangle}~.
\end{align}
Note, that $\omega^X=1$ for the Poisson distribution.
Thus, to study the fluctuations of charged particles
the second moment of the multiplicity distribution 
$\;\langle N_{\pm}^2\rangle\;$
has to
be calculated.
In the $~g.c.e.~$ (\ref{Zgce}) and $c.e.$ (\ref{Zce}) one finds:
\begin{align}\label{N2gce}
\langle N_{\pm}^{2} \rangle_{g.c.e.} &\;=\; \frac{1}{Z_{g.c.e.}}
  \left[ 
\frac{\partial}{\partial \lambda_{\pm}}
  \left( \lambda_{\pm}
        \frac{\partial \;\;Z_{g.c.e.}}{\partial \lambda_{\pm}}
  \right) \right]_{\lambda_{\pm}=1}
  \;=\; z\;+\;z^2~,\\
\langle N_{\pm}^{2} \rangle_{c.e.}
 &\;=\; \frac{1}{Z_{c.e.}}
  \left[ 
\frac{\partial}{\partial \lambda_{\pm}}
  \left( \lambda_{\pm}
        \frac{\partial \;\;Z_{c.e.}}{\partial \lambda_{\pm}}
  \right) \right]_{\lambda_{\pm}=1}
  \;=\; z \;\frac{I_{1}(2z)}{I_{0}(2z)}
  \;+\; z^2 \;\frac{I_{2}(2z)}{I_{0}(2z)}~=~z^2~.\label{N2ce}
\end{align}
The corresponding scaled variances are:
\begin{align}\label{omega-gce}
\omega_{g.c.e.}^{\pm}
 & \;=\; \frac{\langle N_{\pm}^2\rangle_{g.c.e.}
        \;-\; \langle N_{\pm}\rangle_{g.c.e.}^2}
{\langle N_{\pm}\rangle_{g.c.e.}}
   \;=\; 1~,\\
\omega_{c.e.}^{\pm}
 & \;=\; \frac{\langle N_{\pm}^2\rangle_{c.e.}
        \;-\; \langle N_{\pm}\rangle_{c.e.}^2}{\langle N_{\pm}\rangle_{c.e.}}
   \;=\; 1 \;-\; z\left[\,\frac{I_1(2z)}{I_0(2z)}
           \;-\; \frac{I_2(2z)}{I_1(2z)}\,\right]~.\label{omega-ce}
\end{align}
Using Eqs.~(\ref{Bessel-0lim}) and (\ref{Bessel-1lim})
the asymptotic behaviour of $\omega_{c.e}^{\pm}$ for
both $z\rightarrow 0$ and $z\rightarrow \infty$
can be found. 
The c.e. fluctuations measured in terms of $\omega$ are 
equal to those in the $~g.c.e.~$ 
for the small system 
($z\ll
1$) 
(another variable to treat the
fluctuations in the small systems is discussed in Appendix):
\begin{align}\label{omega-ce0}
\omega_{c.e}^{\pm}
  \; \cong \; 1~-~\frac{z^2}{2}\;\cong ~1~ = \omega^{\pm}_{g.c.e}~.
\end{align}
For large
systems $(z\gg 1)$ 
the scaled variance for the $~c.e.~$ 
is two times smaller than the scaled variance for  the
$~g.c.e.~$:
\begin{align}\label{omega-ce1}
        \omega_{c.e.}^{\pm}
  \; \cong \; \frac{1}{2} ~ + ~ \frac{1}{8z}~\cong ~\frac{1}{2}~=~\frac{1}{2}
  ~\omega_{g.c.e.}^{{\pm}}\;.
\end{align}
%
%
The dependence of the scaled variance calculated within
the $c.e$ and $g.c.e.$ on $z$ is shown in Fig.~2.

\newpage
\begin{figure}[h!]
\epsfig{file=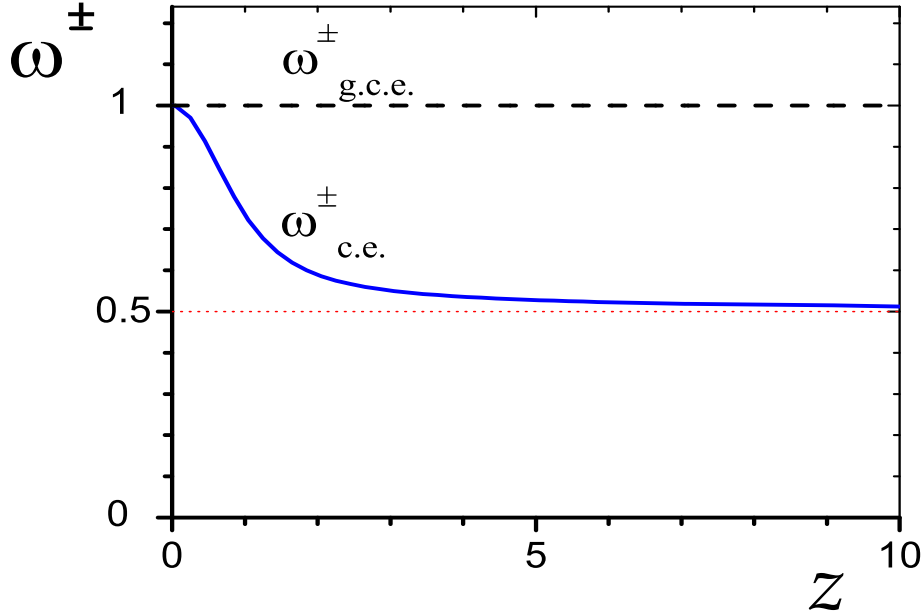,height=10cm,width=14cm}

\vspace{-1cm}
 \caption{ The scaled variances of $N_{\pm}$
calculated within the $~g.c.e.~$, $\omega^{\pm}_{g.c.e.}=1$ 
(\ref{omega-gce}), and $~c.e.~$, $\omega^{\pm}_{c.e.}$
 (\ref{omega-ce}).
} \label{fig-omega}
\end{figure}

\noindent
The scaled variance shows a very different behavior than the mean
multiplicity.
In the limit of small $z$ the ratio of the results for 
$c.e.$ and $g.c.e.$ approaches zero for the mean
multiplicity (Fig. 1) and one for the scaled variance (Fig. 2).
On the other hand in the large $z$ limit the mean multiplicity
ratio approaches one and the scaled variance ratio 0.5.
Thus in the case of fluctuations the canonical and
grand canonical ensembles are not equivalent.

\vspace{0.3cm} \noindent {\bf 5. Multiplicity distribution.}~~ 
In the $g.c.e.$ the multiplicity
distribution of $\;N_+\;$ (and $\;N_-$) is equal  to the Poisson one:
\begin{align}\label{poisson}
P_{g.c.e.}(N_{+})\;& \equiv \; \sum_{N_{-}=0}^{\infty}~
P_{g.c.e.}\left(N_+,N_-\right)~=~
\frac{1}{Z_{g.c.e.}}~ \sum_{N_{-}=0}^{\infty}
\frac{z^{N_{+}}}{N_{+}!}~\frac{z^{N_{-}}}{N_{-}!}\\
~&=~ \exp(-z) \cdot
\frac{z^{N_{+}}}{N_{+}!}~,\nonumber
\end{align}
whereas
the corresponding  distribution  in
the $\;c.e.\;$ (\ref{Zce}) is:
\begin{align}\label{PceN}
P_{c.e.}(N_+)\;& \equiv \;
\sum_{N_{-}=0}^{\infty}~
P_{c.e.}\left(N_+,N_-\right)~=~
\frac{1}{Z_{c.e.}}~\sum_{N_- =0}^{\infty}
\frac{z^{N_{+}}}{N_{+}!}~\frac{z^{N_{-}}}{N_{-}!}~\cdot \delta\left(N_{+}
- N_{-}\right)\\
 \;&=\;
 \frac{1}{I_0(2z)}\cdot\left(\frac{z^{N_+}}{N_+!}\right)^2\;. \nonumber
\end{align}
As an example, the distributions in $g.c.e.$ and $c.e.$ are plotted
in Figs. 3 and 4 for $z = 0.5$ (the small system) and $z = 10$
(the  large system), respectively. 

\newpage
\begin{figure}[h!]
\epsfig{file=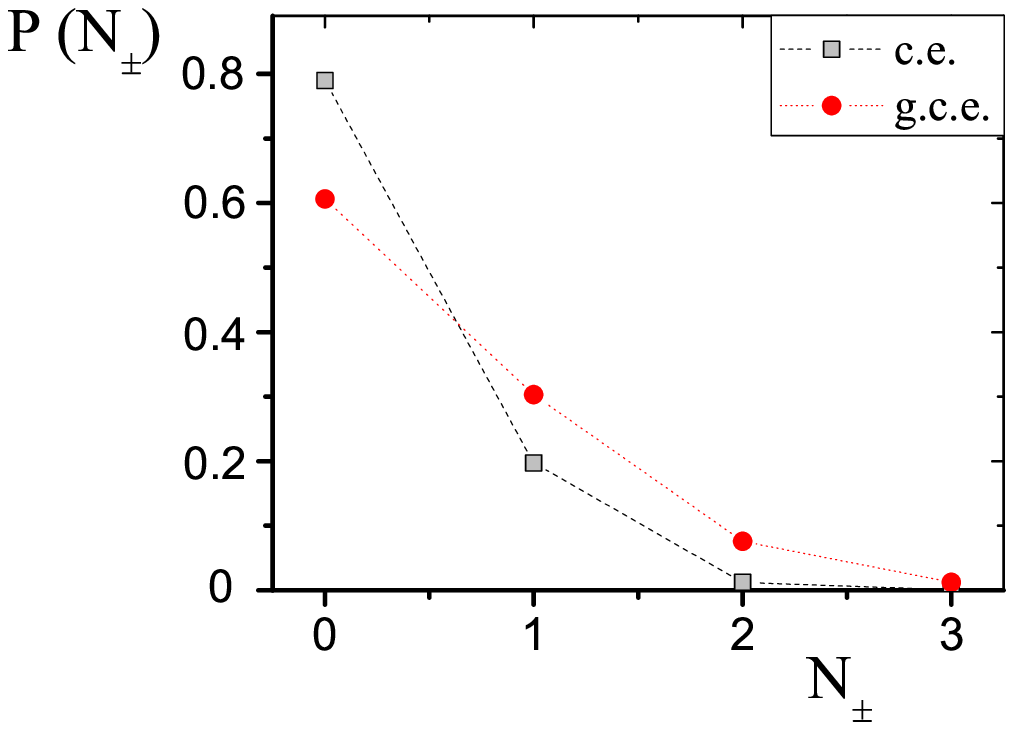,height=10cm,width=14cm}
                                                                                
\vspace{-1cm} \caption{Multiplicity distributions  $P_{c.e.}(N_{\pm})$
(\ref{PceN}) and $P_{g.c.e.}(N_{\pm})$ (\ref{poisson}) for $z=0.5$.}
\label{fig-Pce-05}
\end{figure}

\vspace{-0.5cm}
\begin{figure}[h!]
\epsfig{file=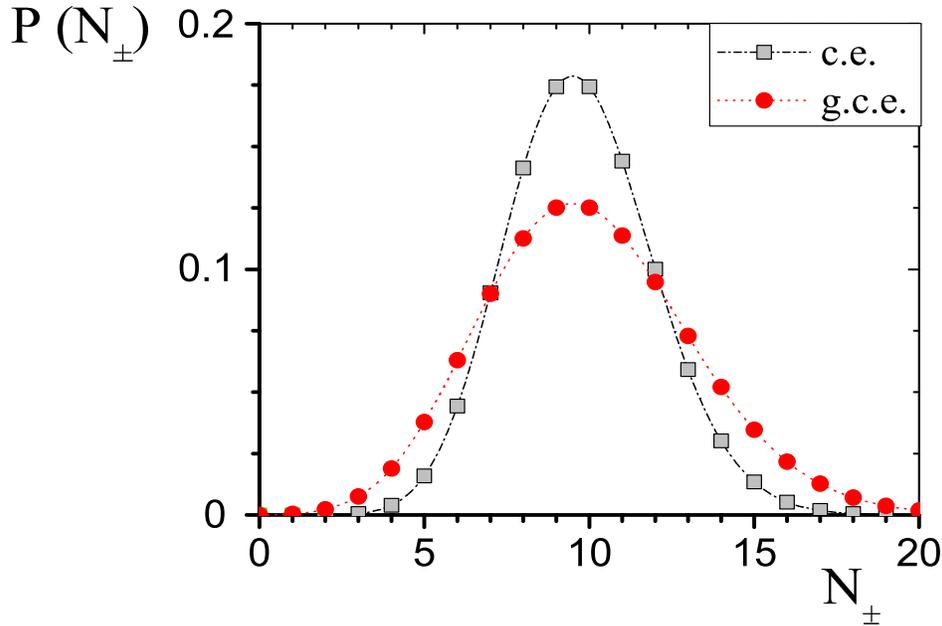,height=10cm,width=14cm}
                                                                                
\vspace{-1cm}
 \caption{Multiplicity distributions
$P_{c.e.}(N_{\pm})$ (\ref{PceN}) and $P_{g.c.e.}(N_{\pm})$ (\ref{poisson})
for $z=10$.
} \label{fig-Pce-10}
\end{figure}

As expected from the previous discussion, the $~c.e.~$ 
distribution (\ref{PceN}) is narrower 
(the variance is smaller)
than  the
$~g.c.e.~$ one (\ref{poisson}). This  result is valid for both
the large ($z\gg 1$) and the small ($z\ll 1$) system. 
On the other hand,
the average value of
$N_{\pm}$ is  smaller in  the $~c.e.~$ 
than in
the $~g.c.e.~$ for small $z$. It results in $\omega^{\pm}_{c.e.}\rightarrow
\omega^{\pm}_{g.c.e}=1$ at $z\rightarrow 0$. Moreover, for $\langle
N_{\pm} \rangle \ll 1$ one can easily demonstrate that $\omega^{\pm} \cong
1$ for any $P(N_{\pm})$ distribution if the conditions $P(0)\gg P(1)\gg
P(k)$ (with $k\ge 2$) are satisfied. Indeed, in this limit  one can
neglect all $P(N_{\pm})$ for
$N_{\pm}\ge 2$ 
which results in:
\begin{align}\label{small-z}
\omega^{\pm}
 \; \equiv \;\frac{\langle N_{\pm}^2\rangle - \langle
N_{\pm}\rangle^2}{\langle N_{\pm}\rangle} \; \cong \;
 \frac{P(1)\cdot 1^2 \;-\; [P(1)\cdot 1]^2}
      {P(1)\cdot 1} 
\; \cong~ 1~,
\end{align}
as $P(1)\cong \langle N_{\pm}\rangle \ll  1$.
In the large volume limit, see Fig. 4, 
the mean values of the  $c.e.$ and $g.c.e.$ distributions
become equal, but the $c.e.$ distribution is narrower 
than the $g.c.e.$ one.

\begin{figure}[h!]
\vspace{-1cm}
 \epsfig{file=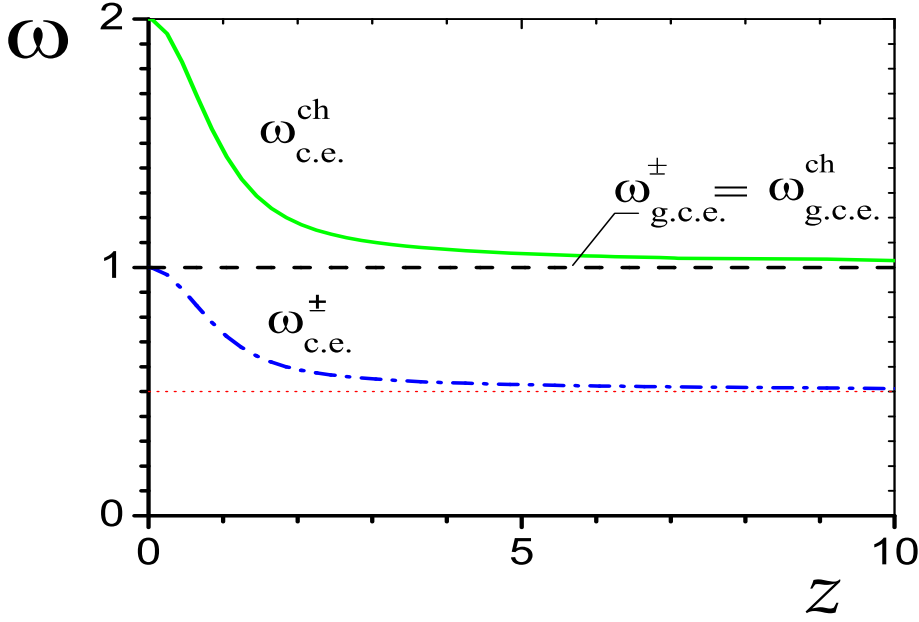,height=10cm,width=14cm}
\vspace{-1cm} \caption{The scaled 
variances $\omega^{ch}_{c.e.}$~
(\ref{omega-ch}),~~ $\omega^{\pm}_{c.e.}$~ (\ref{omega-ce})~~ and
$\omega^{\pm}_{g.c.e.} =\omega^{ch}_{g.c.e.}=1$~
(\ref{omega-gce},\ref{omega-ch-gce})
as functions of $z$. 
}\label{omega-all}
\end{figure}

\vspace{0.3cm} 
\noindent
{\bf 6. Total multiplicity of charged particles.}~~ The total 
multiplicity of charged particles is defined as
$N_{ch} = N_++N_-$. Its average  in the $\;g.c.e.\;$
and $~c.e.~$ reads:
\begin{align}\label{Nch-gce}
\langle N_{ch} \rangle_{g.c.e.}
 &\;=\; \langle\;N_+ + N_-\;\rangle_{g.c.e.}
 \;=\; \langle N_+\rangle_{g.c.e.} + \langle N_-\rangle_{g.c.e.}
 \;=
 ~2z\;,\\
\label{Nch-ce}
\langle N_{ch} \rangle_{c.e.}
 &\;=\; \langle\;N_+ + N_-\;\rangle_{c.e.}
 \;=\; \langle N_+\rangle_{c.e.} + \langle N_-\rangle_{c.e.}
 \;=\; 2z\;\frac{I_1(2z)}{I_0(2z)}\;.
\end{align}
In the $~g.c.e.~$ one finds:
\begin{align}\label{Nch2gce}
 \langle N_{ch}^{2} \rangle_{g.c.e.}&~=~\langle N_{+}^{2}
+2N_{+}N_{-}+N_{-}^{2}\rangle_{g.c.e.} ~=~\langle
N_{+}^{2}\rangle_{g.c.e.}+2\langle N_{+}\rangle_{g.c.e.}\langle
N_{-}\rangle_{g.c.e.}+\\
&~+~\langle N_{-}^{2}\rangle_{g.c.e.}
~=~z^{2}+z~+~2z^{2}~+~z^{2}+z~=~4z^{2}~+~2z~,\nonumber
\end{align}
and consequently the scaled variance of $N_{ch}$ in the
$~g.c.e.~$ is:
\begin{align}\label{omega-ch-gce}
\omega_{g.c.e.}^{ch}~\equiv~\frac{\langle
N_{ch}^{2}\rangle_{g.c.e.}~-~\langle N_{ch}\rangle^{2}_{g.c.e.}}{\langle
N_{ch}\rangle_{g.c.e.}}~=~ \frac{4z^{2}+2z~-~(2z)^{2}}{2z}~=~1~.
\end{align}
The result (\ref{omega-ch-gce}) also  follows  
from explicit expression on the probability
distribution of $N_{ch}$ in the $~g.c.e.~$:
\begin{align}\label{Pch-gce}
&P_{g.c.e.}(N_{ch})~\equiv~
\sum_{N_{+}}^{\infty}
\sum_{N_{-}=0}^{\infty}
P_{g.c.e.}\left(N_+,N_-\right)
\cdot \delta\left[N_{ch}-\left(N_{+}+N_{-}\right)\right]\\
&~=~
\frac{1}{Z_{g.c.e.}}~\sum_{N_{+}}^{\infty}
\sum_{N_{-}=0}^{\infty}
\frac{z^{N_{+}}}{N_{+}!}~\frac{z^{N_{-}}}{N_{-}!}\cdot
\delta\left[N_{ch}-\left(N_{+}+N_{-}\right)\right]
~= ~\exp(-2z)~\frac{(2z)^{N_{ch}}}{N_{ch}!}~.\nonumber
\end{align}
Thus
distributions of $N_{ch}$ and $N_{\pm}$ are Poissonian in the
$~g.c.e.~$. 
In the $~c.e.~$ the negatively and positively charged particles are
correlated,
$\langle N_+\cdot N_-\rangle_{c.e.}
 \neq \langle N_+\rangle_{c.e.}\cdot\langle N_-\rangle_{c.e.}$.
The correlation term reads:
\begin{align}\label{N+N-}
\langle N_+\cdot N_-\rangle_{c.e.}
 \;=\; \frac{1}{Z_{c.e.}}
    \left( 
    \frac{\partial^2\;Z_{c.e.}}{\partial \lambda_+\partial \lambda_-}
  \right)_{\lambda_{\pm}=1}~
    =\; z^2~.
\end{align}
Using Eqs.~(\ref{N2ce}) and (\ref{N+N-}) one obtains the scaled variance of
$N_{ch}$ in the $~c.e.~$:
\begin{align}\label{omega-ch}
 \omega_{c.e.}^{ch}
  \;\equiv ~ \frac{\langle N_{ch}^2\rangle_{c.e.}
        \;-\; \langle N_{ch}\rangle^2_{c.e.}}{\langle N_{ch}\rangle_{c.e.}}
   \;=\; 1 \;+\; z \left[ \frac{I_2(2z) + I_0(2z)}
                                {I_1(2z)}
       \;-\; 2\;\frac{I_1(2z)}{I_0(2z)} \right]~.
\end{align}
%
The scaled variances $\omega^{ch}_{g.c.e}$
and $\omega^{ch}_{c.e.}$
as functions of $z$ are shown in Fig.~\ref{omega-all} together with
$\omega^{\pm}_{g.c.e}$
and $\omega^{\pm}_{c.e.}$.

\newpage

From Eqs.~(\ref{omega-ce}) and (\ref{omega-ch}) and the recurrence
relation
$I_0(2z)=I_2(2z)+I_1(2z)/z$ \cite{I} it follows
that $\omega_{c.e.}^{ch}= 2 \omega_{c.e.}^{\pm}$, i.e.
the relative variance of total charge multiplicity $N_{ch}$ is two times
larger than the one of $N_{\pm}$. This is because $N_{ch}=2N_{+}=2N_{-}$
in each microscopic state allowed by an exact charge conservation. 
One obtains a similar result for 
the case of particle production via decay of neutral
resonances,
e.g., $\rho^0\rightarrow\pi^+ + \pi^-$. 
The distributions of $\pi^{+ }$ and $\pi^{-}$ coincide with
the $\rho^{0}$ distribution, and consequently $\omega^{\pm}=\omega$, where
$\omega$ is the scaled variance of the
distribution of $\rho^{0}$. 
But because 
$N_{ch} = 2 N_{\rho}$ one gets 
$\omega^{ch}=2\omega$.

Probability distribution of $\;N_{ch}\;$ in the $~c.e.~$ 
reads:
\begin{align}\label{Pch-ce}
P_{c.e.}(N_{ch})
&\; \equiv\;
\sum_{N_{+}}^{\infty}
\sum_{N_{-}=0}^{\infty}
P_{c.e.}\left(N_+,N_-\right)
\cdot \delta\left[N_{ch}-\left(N_{+}+N_{-}\right)\right]\\
&~=~
\frac{1}{I_{0}(2z)}~\sum_{N_+=0}^{\infty}\sum_{N_-=0}^{\infty}
\frac{z^{N_{+}}}{N_{+}!}\frac{z^{N_{-}}}{N_{-}!}\cdot
\delta\left(N_{+}-N_{-}\right) \cdot
\delta\left[N_{ch}-\left(N_++N_-\right)\right] \nonumber \\ 
&\;=\; \frac{1}{I_0(2z)}~                 
 \left[\frac{z^{N_{ch}/2}}{(N_{ch}/2)!\;}\right]^2\;.\nonumber
\end{align}
It coincides, of course, with $P_{c.e.}(N_{+})$ (\ref{PceN}) at
$N_{+}=N_{ch}/2$. As an example , the probability
distributions $P_{g.c.e.}(N_{ch})$ (\ref{Pch-gce}) and $P_{c.e.}(N_{ch})$
(\ref{Pch-ce}) are shown for  $z=0.5$ (the small system)
and for $z =10$ (the large system) in Figs. 6 and 7, respectively.
Only even multiplicities $N_{ch}=0,\;2,\;4\ldots\;$
are allowed in the $~c.e.~$ because of an exact charge conservation.
For the small system ($z\ll 1$) the $\omega^{ch}$ reads
(both $P_{g.c.e.}(N_{ch}=1)\ll 1$ and $P_{c.e.}(N_{ch}=2)\ll 1$ at
$z\ll 1$):
\begin{align}\label{omega-ch1}
\omega^{ch}_{g.c.e.} & ~ \cong ~
 \frac{P_{g.c.e.}(1)\cdot 1^2
       ~ - ~ \left[P_{g.c.e.}(1)\cdot 1\right]^2}
      {P_{g.c.e.}(1)\cdot 1}
~ \cong ~ 1~,\\
\omega^{ch}_{c.e.}
 & ~ \cong ~
 \frac{P_{c.e.}(2)\cdot 2^2
      \;-\; \left[P_{c.e.}(2)\cdot 2\right]^2
      \;-\; \left[P_{c.e.}(2)\cdot 2\right]^2}
      {P_{c.e.}(2)\cdot 2}
~\cong ~2~.\label{omega-ch2}
\end{align}


\begin{figure}[h!]
\epsfig{file=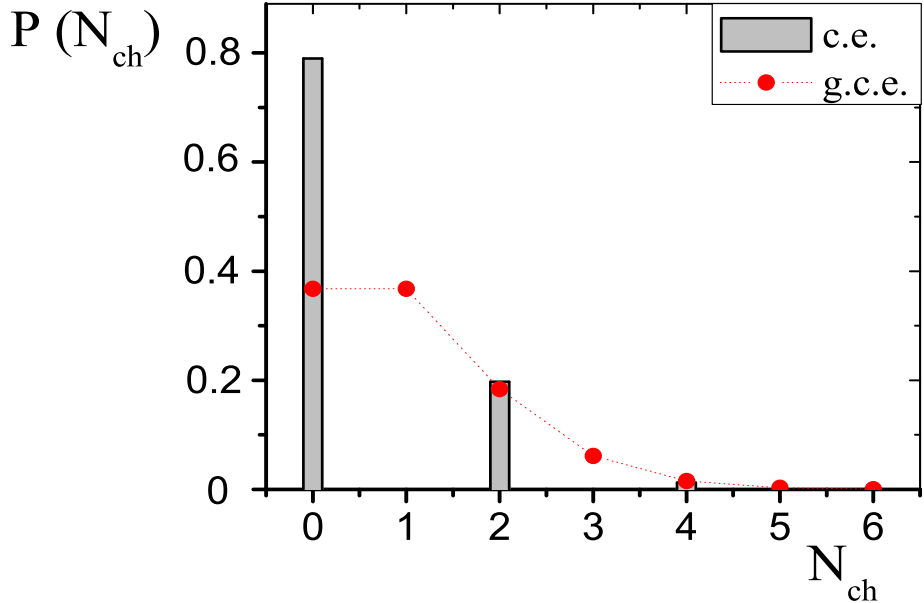,height=10cm,width=14cm}

\vspace{-1cm}
 \caption{Multiplicity  distributions of $\;N_{ch}\;$ for
$\;z=0.5\;$ in the $\;g.c.e.\;$ and $\;c.e.\;$.
}
\label{fig-Pce-ch-05}
\end{figure}
%
%
\begin{figure}[h!]
\epsfig{file=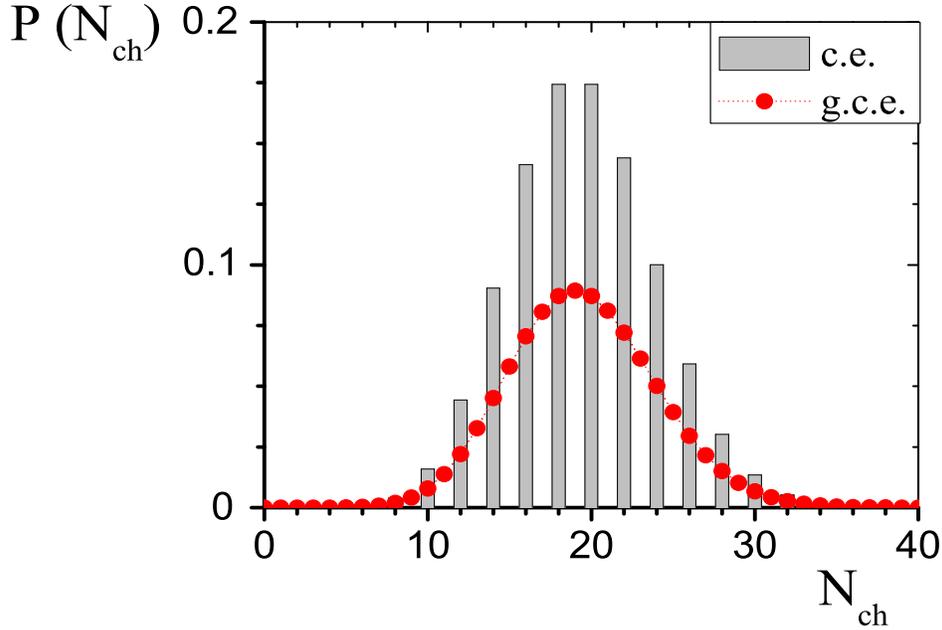,height=10cm,width=14cm}
                                                                                
\vspace{-1cm}
 \caption{Multiplicity  distributions of $\;N_{ch}\;$ for
$\;z=10\;$ in the $\;g.c.e.\;$ and $\;c.e.\;$.}
\label{fig-Pce-ch-10}
\end{figure}

In the large $z$ limit  the average number of charge
particles $\langle N_{ch} \rangle$ and its scaled variance $\omega^{ch}$
 in the $\;g.c.e.\;$, Eqs.~(\ref{Nch-gce}) and (\ref{omega-ch-gce}),
are equal to those
in the $\;c.e.\;$, Eqs.~(\ref{Nch-ce}) and (\ref{omega-ch}).
Nevertheless
the corresponding probability distributions
are different, see Fig. 7. 
This is because all odd multiplicities are excluded in
$c.e.$ as a consequence of the charge conservation.
The relation between
$P_{g.c.e.}(N_{ch})$ (\ref{Pch-gce}) and $P_{c.e.}(N_{ch})$
(\ref{Pch-ce})  for the large system ($z\gg
1$) 
can be established as follows.
Let us introduce the probability distribution $P^{*}(N_{ch})$
defined as
\begin{align}\label{P*}
P^{*}(N_{ch})~&\equiv~
C\cdot P_{g.c.e.}(N_{ch})~,&~~N_{ch}&\;=\;0,\;2,\;4,\ldots~,\\
P^{*}(N_{ch})~&\equiv~ 0~,&~~N_{ch}&\;=\;1,\;3,\;5,\ldots~,
\end{align}
where the constant $C$ is given by a normalization condition
\begin{align}\label{C}
1&~=~\sum_{N_{ch}=0,2,4,\ldots}~P^{*}(N_{ch})~\equiv~ C\cdot
\sum_{N_{ch}=0,2,4,\ldots}~P_{g.c.e.}(N_{ch})\\
&~=~C\cdot \exp(-2z)~\sum_{n=0}^{\infty} \frac{(2z)^{2n}}{(2n)!}~=~C\cdot
\exp(-2z)~\cosh(2z)~.\nonumber
\end{align}
Using Eq.~(\ref{C}) one gets $C=2\cdot [1+\exp(-4z)]^{-1}
\cong 2$ for $z\gg 1$. The origin of the result $C\cong 2$ is the
fact that
\begin{align}
P_{g.c.e.}(N_{ch}+1)~\equiv~ P_{g.c.e.}(N_{ch})\cdot
\frac{2z}{N_{ch}+1}\cong P_{g.c.e.}(N_{ch})~,
\end{align}
for $N_{ch}$ close to its average value $\langle
N_{ch}\rangle_{g.c.e.}= 2z\gg 1$, i.e. if the odd numbers
$N_{ch}=1,3,5,\ldots~$ are forbidden the probabilities
$P_{g.c.e.}(N_{ch})$ for the even numbers $N_{ch}=0,2,4,\ldots~$
should be approximately doubled to have a correct normalization
for $P^{*}(N_{ch})$ (\ref{P*}). 

\vspace{-0.5cm}
\begin{figure}[h!]
\epsfig{file=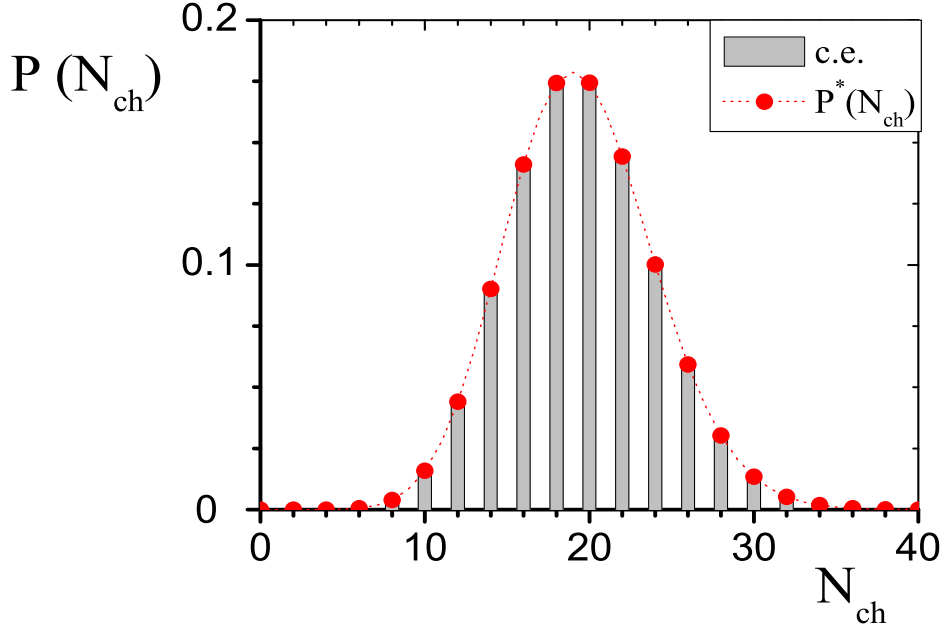,height=10cm,width=14cm}

\vspace{-1cm}
 \caption{Multiplicity distributions
 $P_{c.e.}(N_{ch})$ (\ref{Pch-ce}) and
  $P^{*}(N_{ch})$ (\ref{P*})
for $z=10$.
}\label{PceNch_10*}
\end{figure}

 Using the Stirling
formula, $n!\cong n^ne^{-n}\sqrt{2\pi n}$, valid for $n\gg 1$, one finds
that $P_{c.e.}(N_{ch})\cong P^{*}(N_{ch})$ for $N_{ch}$ close to its
average value equal to $2z\gg 1$.
Both distributions are plotted in Fig. 8 for a comparison.

\vspace{0.3cm}
\noindent
{\bf 7. Limited kinematical acceptance.}~~
In the experimental study of nuclear collisions at high energies
only a fraction of all produced particles which carry conserved 
charges is registered.
Thus the multiplicity distribution of the measured particles is expected to
be different from the distribution of all produced particles.
Within $c.e.$ the effect of the limited kinematical acceptance
(the acceptance in the momentum space) can be taken into 
account introducing a probability $q$ that a single particle
is registered.
Because in $c.e.$ particles are uncorrelated in momentum space
the multiplicity distribution of accepted particles for a fixed 
number of produced particles $N$ is
given by the binomial distribution:
\begin{align}\label{bin}
P_{acc}(n,N)~=~q^n(1-q)^{N-n}~\cdot \frac{N!}{n!(N-n)!}~.
\end{align}
Consequently one gets:
\begin{align}\label{av}
\overline{n}~=~q~N~,~~~~~
\overline{n^2}~-\overline{n}^2~=~q(1-q)~N~,
\end{align}
where $(k=1,2)$
\begin{align}\label{av1}
\overline{n^k}~\equiv~\sum_{n=0}^{N} n^k~ P_{acc}(n,N)~.
\end{align}
Introducing the probability distribution $P(N)$ the first two 
moments of the distribution of accepted particles can be 
calculated:
\begin{align}\label{ac1}
\langle n \rangle_{acc} & ~\equiv~\sum
_{N=0}^{\infty}P(N)~ \sum_{n=0}^{N}
n~P_{acc}(n_+,N)~=~q \cdot \langle N\rangle~,\\
\langle n^2 \rangle_{acc} & ~\equiv~\sum
_{N=0}^{\infty}P(N)~ \sum_{n=0}^{N}
n^2~P_{acc}(n,N)~=~q^2\cdot\langle N^2\rangle~
+~q (1-q) \cdot \langle N \rangle ~,\label{ac2}
\end{align}
where ($k=1, 2$)
\begin{align}\label{ac3}
\langle N^k \rangle ~ \equiv~ \sum_{N=0}^{\infty} 
N^k~P(N)~.
\end{align}
Finally, the scaled variance for the accepted particles can be
obtained:
\begin{align}\label{ac4}
\omega_{acc}~=~q\cdot \omega~ +~(1-q)~,
\end{align}
where $\omega$ in Eq.~(\ref{ac4}) is the scaled variance of the $P(N)$
distribution. Assuming that $P(N)$ corresponds to the $~c.e.~$,
one finds from Eq.~(\ref{ac4}) 
the scaled variance for the accepted
particles in the $~c.e.~$
$\omega_{acc}^+\cong 1$ for $q\ll 1$ and $\omega_{acc}^+\cong
\omega_{c.e.}^{+}$  for $q\cong 1$.

These limiting behaviour agrees with the expectations.
In the large acceptance limit ($q \approx 1$) the distribution of
measured particles approaches the distribution in the full
acceptance. 
For a very small acceptance ($q \approx 0$) the measured distribution
approaches the Poisson one independent of the shape of the
distribution in the full acceptance.

\vspace{0.3cm}
\noindent { \bf 8. Summary.} ~~
The particle number fluctuations have been considered within
canonical ensemble for a system with zero net charge.
The results are compared to those
in the grand canonical ensemble where only the mean value
of charge is required to be zero.
In the large volume limit the fluctuations in $c.e.$
are found to be different from those in the $g.c.e.$ .
Thus the well known  equivalence of both ensembles
for the mean quantities is not valid for the fluctuations.
The scaled variance of the multiplicity distribution of
same charge particles is  calculated to be 0.5 in the $c.e.$  and
it is two times smaller than the scaled variance in $g.c.e$ .
These results may be relevant for the analysis of fluctuations
in high energy nuclear collisions.
In view of this the influence of the limited kinematical acceptance
on multiplicity fluctuations also have been discussed.

In this work the influence of the  electric charge conservation was discussed.
However, other material conservation laws, e.g. baryon number,
strangeness or charm, can be treated within the same scheme.  
An extension of this work for a non-zero
value of the conserved charge and  several  species of charged
particles as well as  an influence of an exact charge conservation on the energy
fluctuations in the $~c.e.~$ will be presented elsewhere.

\vspace{0.3cm}
\noindent
{\bf Acknowledgments.}~~We are grateful to A.I.~Bugrij, A.P.~Kostyuk,
I.N.~Mishustin, L.M.~Satarov and Yu.M.~Sinyukov for critical comments and
useful discussions. 
We thank Marysia Gazdzicka for help in preparing the manuscript.
Partial support by Institut f\"ur Theoretische Physik,
Frankfurt Universit\"at and  Frankfurt Institute for Advanced Studies
(M.I.G.), by DAAD scholarship under Leonard-Euler-Stipendienprogramm
(O.S.Z.) and by
Polish Committee of Scientific Research under grant 2P03B04123 (M.G.) 
is acknowledged.

\begin{figure}[h!]
\epsfig{file=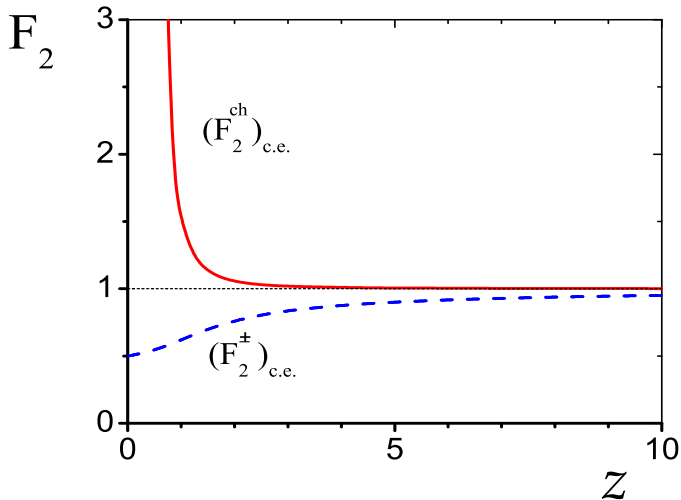,height=8cm,width=10cm}
                                                                                
\vspace{-1cm}
 \caption{The fluctuation measure $F_{2}$ as a function
of $z$  in the
$~c.e.~$. The dashed and solid lines indicate $\left(F_2^{\pm}\right)_{c.e.}$
(\ref{F2pm}) and 
$\left(F_2^{ch}\right)_{c.e.}$ (\ref{F2ch}), respectively.}\label{F2}.
\end{figure}

\vspace{0.5cm}
\noindent
{\bf Appendix.}~~
A variable
$F_2^X~=~\langle X(X-1) \rangle / \langle X\rangle^2$~,
was used for study of the fluctuations of $N_{\pm}$ and $N_{ch}$
in small systems ($z\ll 1$)
\cite{koch}. 
In the $~g.c.e.~$ (i.e. for the Poisson distribution of $N_{\pm}$
and $N_{ch}$) one gets
$\left(F_{2}^{\pm}\right)_{g.c.e.}=\left(F_{2}^{ch}\right)_{g.c.e.}=1$,
whereas in the $~c.e.~$ one finds (see also Fig.~(\ref{F2})):                                 
\begin{align}\label{F2pm}
\left(F_{2}^{\pm}\right)_{c.e.} &~\equiv ~\frac{\langle N_{\pm}
~\left(N_{\pm}-1 \right)\rangle_{c.e.}}{\langle
N_{\pm}\rangle^{2}_{c.e.}}~=~
\frac{I_{0}(2z)\cdot I_{2}(2z)}{I_{1}^2(2z)}~~
\raisebox{-.5ex}{$\stackrel{ \cong}{\scriptstyle z\ll 1}$}~~
\frac{1}{2}~+~\frac{z^2}{6}~,\\
\left(F_{2}^{ch}\right)_{c.e.} &~\equiv ~\frac{\langle N_{ch}
~\left(N_{ch}-1 \right)\rangle_{c.e.}}{\langle
N_{ch}\rangle^{2}_{c.e.}}~=~
\frac{I_{0}^2(2z)~+~I_0(2z)\cdot I_{2}(2z)}{2~I_{1}^2(2z)}~~
\raisebox{-.5ex}{$\stackrel{ \cong}{\scriptstyle z\ll 1}$}~~
\frac{1}{2z^2}~.\label{F2ch}
\end{align}

\vspace{-0.5cm}

\vspace{-0.3cm}
\noindent
In the large volume limit   
$F_2^{\pm} \rightarrow 1$
and $F_2^{ch} \rightarrow 1$
because  $\langle N_{\pm}\rangle^{2} \gg
\langle N_{\pm}\rangle \gg 1$ (the same for $N_{ch}$).
Therefore, this measure is not suitable for a study of
the particle number fluctuations in the large systems.

\end{document}